\documentclass[twocolumn,10pt,final,conference]{IEEEtran}

\usepackage{cite}
\usepackage{amsmath}
\usepackage{amssymb}
\usepackage{graphicx,dblfloatfix}
\usepackage{subfigure}
\usepackage{a4wide}
\usepackage{algorithm}
\usepackage{algorithmic}
\usepackage{amsfonts}
\usepackage{verbatim}

\makeatletter
\def\ps@headings{%
\def\@oddhead{\mbox{}\scriptsize\rightmark \hfil \thepage}%
\def\@evenhead{\scriptsize\thepage \hfil \leftmark\mbox{}}%
\def\@oddfoot{}%
\def\@evenfoot{}}
\makeatother
\begin{document}
\title{Analyzing Cascading Failures in Smart Grids under Random and Targeted Attacks}

\author{\IEEEauthorblockN{
Sushmita Ruj$^\dag$ and Arindam Pal$^\ddag$}
\IEEEauthorblockA{
$^\dag$ Indian Statistical Institute, Kolkata, India. Email: sush@isical.ac.in \\
$^\ddag$ TCS Innovation Labs, Kolkata, India. Email: arindamp@gmail.com
}}

\maketitle{}

\begin{abstract}
We model smart grids as complex interdependent networks, and study targeted attacks on smart grids for the first time. 
A smart grid consists of two networks: the power network and the communication network, interconnected by edges.
Occurrence of  failures (attacks) in one network triggers failures in the other network, and propagates in cascades across the networks. 
Such cascading failures can result in disintegration of either (or both) of the networks. 
Earlier works considered only random failures. 
In practical situations, an attacker is more likely to compromise nodes selectively.  

We study cascading failures in smart grids, where an attacker selectively compromises the nodes with probabilities proportional to their degrees; high degree nodes are
compromised with higher probability. 
We mathematically analyze the sizes of the giant components of the networks under targeted attacks, and compare the results with the corresponding sizes under random attacks.
We show that networks disintegrate faster for targeted attacks compared to random attacks. 
A targeted attack on a small fraction of high degree nodes disintegrates one or both of the networks, 
whereas both the networks contain giant components for random attack on the same fraction of nodes. 
\end{abstract}


\section{Introduction}
\label{sec:intro}
Smart grids are next generation electricity grids, in which the power network and the communication network work in tandem.
Power grids have suffered severe attacks in the past. 
The black out of Northern US/Canada and that of Italy in 2003 affected the lives of millions of people and resulted 
in huge monetary losses. 
More recently, the largest blackout in the world occurred in India in July 2012.
The complete shutdown of the  Northern, Eastern, and Northeastern power grids in the country affected over 620 million people.
Such calamities could have been avoided, if the power grid functioned properly. 
In order to ensure that the electric grid functions smoothly, it is important that the control
information is collected and transmitted in an orderly fashion, and the existing systems be highly automated. 
Smart grids promise to fulfill this vision by synchronizing the power network with the communication network. The idea is to replace the existing \emph{SCADA (Supervisory Control and Data Acquisition)} system  by an intelligent and automatic communication network. 

The power network consists of power plants, generation and distribution stations, whereas the communication network
consists of sensors attached to appliances to collect information, aggregator sensors to aggregate information and smart meters for monitoring and billing. 
The smart meters in home area networks, building area networks, and
neighborhood area networks are responsible for aggregating, processing and transmitting data and control information for proper functioning of the smart grid. 
The question is how to make such a network robust and fault-tolerant. 
Researchers have addressed smart grid architectures \cite{B10} and the problem of cascading failures \cite{CPS11}, 
in which a small fault propagates throughout the network and affects a large part of the network. 
Most of the current techniques and models use concepts from distributed systems. 
However, because of the large size of smart grids and their unique properties, new models, 
interconnection patterns, and analysis techniques are required to increase  the robustness of networks.

Recently, Huang \emph{et al.} \cite{HWRSN13} initiated the study of modeling and analyzing smart grids using
interdependent  complex networks.   
A smart grid can be thought of as two complex networks, which are interconnected. 
The question is how to make this network robust and fault tolerant. 
In order to provide a solution, we have to understand what kind of faults and attacks can take place and 
how faults propagate in the network.  
The failure of nodes in one network results in the disruption of the other network, which in turn affects the first network. 
This type of failure propagates in a cascading manner and was the main reason for the blackouts in the US and in India. 
To understand this \emph{cascading failure}, we need to study the structure of the networks.
In this paper, we model and study smart grids as complex networks and show the effect of cascading failure, when adversaries compromise nodes in the network. 

Though cyber-security issues have been studied in details \cite{WL13}, modeling the network
in order to make it resilient still needs lot of research.  
The main contribution of this paper is to study the effect of targeted attacks in smart grids, 
in which the attacker selectively disrupts communication nodes. 
To the best of our knowledge, this is the first work on targeted attacks on smart grids using complex network model.
We argue that an adversary is more likely to attack selected high degree nodes, rather than attacking nodes randomly. 
As an example, we consider the recent Stuxnet worm \cite{STUX13} which was targeted on Siemens PCs and caused large-scale destruction to industrial control systems.  
Yagan \emph{et al.} \cite{YQZC12} studied cascading failures in cyber-physical systems. 
They studied different interdependent Erdos-Renyi (ER) networks \cite{N10}, but they did not consider scale-free networks, which are used to 
represent power and communication networks. 
Till date, all works \cite{HWRSN13,HWSN13,YQZC12} on complex networks models of smart grid have considered only random attacks.
Huang \emph{et al.} \cite{HWSN13} addressed the cost of maintaining such networks by analyzing the number of support links between networks.
Whereas increasing the support links might make the interdependent networks stronger, large number of support links
imply higher cost of maintenance. They suggested that smart grids should have some nodes which are connected to power nodes (also called \emph{operation centers})
and the rest of the nodes are \emph{relaying nodes}. Using such a model, they studied the resilience of the network under random attacks.
According to their model, each control node is linked to $n$ power nodes and each power node is operated by $k$ operation centers.


\subsection{Problem statement and our contribution}
We model the smart grid as a complex interdependent network consisting of two networks, the power network and the communication network. 
Both the power network and the communication network are scale-free (SF) networks, where the degree distribution follows the power law, 
$p_k \propto k^{-\alpha}$, where $p_k$ is the fraction of nodes of degree $k$ and $\alpha$ is the power-law parameter specific to the network. 
Support links are randomly assigned from one network to another, such that a power node is controlled by multiple communication nodes, and
functions properly as long as at least one such link exists. 
In our model, we consider targeted attacks on the communication network. 
We mathematically analyze the effect of cascading failure for this type of attack and 
find out the sizes of giant components when nodes are compromised. 

We compare the following attack models: random attacks, targeted attacks, and a combination of targeted and random attacks.
We show that 
an adversary has a definite advantage if it compromises nodes selectively. 
In targeted attack, the adversary compromises a node with a probability proportional to the degree of the node. 
Our main conclusion is that by launching a targeted attack, an adversary can disrupt significant part of the network. 
For a large network, compromising about 2.2\% of the network can disrupt either of the networks under targeted attack, whereas under random attack, the networks are still connected and work smoothly. 

\subsection{Organization}
The paper is organized as follows: 
Related works are presented in Section \ref{sec:related}. 
The network model, attack model, and preliminary material on complex networks is presented in Section \ref{sec:model}. 
Cascading failure is mathematically analyzed in Section \ref{sec:analysis}. 
In Section \ref{sec:experimental}, we present experimental results to understand our model and make some conclusions. 
We conclude in Section \ref{sec:conclusion} with directions for future work.

\section{Related works}
\label{sec:related}
Smart grid communication and network architecture have been widely studied in \cite{B10,WXK11,LXLLC12}.
Most smart grid literature concentrate on distribution of power \cite{YLR12}, balancing supply and demand \cite{RWJSL10}, 
detecting and predicting faults \cite{CPS11}, designing network architecture which are fault tolerant \cite{YQZC12}. 
The bulk of literature on fault tolerance address cyber-physical systems in general \cite{YQZC12} and use general models and techniques 
of distributed systems. 

Fault tolerance in power grids has been studied widely in the past. 
The study of the model, analysis of structure, increasing the robustness
of power grids have been studied using complex networks. 
Here, electric distribution stations, 
transmission stations, generation centers are modeled as nodes. Two nodes are connected by a link, if there is power flow from
one node to the other. The structure of the underlying graph has been widely studied, to find the effect of node failures. 
When certain nodes fail (or are attacked), the corresponding links are disrupted. This affects other nodes, whose links fail in return. 
Such failures propagate in a cascading manner throughout the network. Thus, a small fraction of nodes can disrupt a large part of the network.
It has been shown that the graph structure underlying a power grid follows a power law distribution \cite{N10}. 
An extensive survey appears in \cite{PA13}. 
 
Although, complex networks have been widely used to study different networks like social networks, biological networks, citation networks, 
power networks, etc \cite{N10}, smart grid networks have not been widely studied. 
Huang \emph{et al.} \cite{HWRSN13} introduced the study of smart grids using complex interdependent networks, in which the power network and the communication network are modeled as individual networks which have scale-free property. 
The links connecting nodes within a network are called \emph{intralinks}. 
The networks are connected to each other via links (also called \emph{interlinks}), 
such that a power node depends on communication nodes and vice versa. 
Such a network is called \emph{interdependent network}. 

Interdependent networks were introduced by Buldyrev \emph{et al.} \cite{BPPSH10}. They studied the effect of failure cascades in such networks.
The failure of a few nodes in the communication network will affect nodes in the power network, which will further affect nodes in the communication network.
Thus, failures propagate in cascades till a steady state is reached or when either or both of the  networks disintegrate. 
We say that a network \emph{disintegrates} if there are no \emph{giant components} in the network. A giant component is a connected component of size $\Theta(N)$, where $N$ is the number of nodes in the network. Since then, a number of researchers have analyzed interdependent networks. 

The initial study by Buldyrev \emph{et al.} \cite{BPPSH10} studied the case where the two networks are of the same size, and there is a one-to-one correspondence between nodes which are joined by an interlink. 
Shao \emph{et al.} \cite{SBHS11} studied multiple support interlinks, where a node in the power network was connected to multiple nodes in the communication network and vice-versa. 
Most of the results have been analyzed experimentally, because closed-form analytical solutions are difficult to obtain. 
A special case of support links, where nodes having identical degree are connected across networks was studied  in \cite{BSC11}. 
It has been observed in all these cases that interdependent systems make the network much more vulnerable to attacks, compared to a single network. 

A well-known result in complex networks is that, randomly removing 95\% of the nodes in the Internet (which is a scale-free network) can still result in a connected network. However, strategically removing even 2.5\% of the nodes can disrupt the whole network \cite{D07}. 
Such a result motivates us to study the effect of targeted attacks on smart grids. 
In case of smart grids, an adversary is more likely to compromise nodes of strategic importance like hubs, than nodes of low degree. 
Thus, selective attacks give substantially different results compared to random attacks. 

Targeted attacks on interdependent networks has been studied in \cite{HGBHS11}. 
They considered networks of the same size, each node in the power network being connected to an unique node in the communication network. 
The nodes are attacked such that a high degree node has higher probability of being attacked. 
They studied different networks like SF-SF or ER-ER and experimentally calculated the critical probabilities at which networks disintegrate
They showed that targeted attack leads to faster disintegration than random attack. In other words, 
strategically compromising a few nodes results in the removal of the giant component, where as compromising even a large number of nodes randomly, does not result in the removal the giant component. 

Instead of studying interdependent networks consisting of two networks, Dong \emph{et al.} \cite{DGDTSH13} studied targeted attacks on a network of networks. 
Zheng and Liu \cite{ZL12} proposed a solution for making a network robust against targeted attacks by suggesting a onion-like structure. 
Here high-degree nodes are present towards the center in clusters and low-degree nodes are present in concentric rings depending upon their degree. 
They analyzed results from power networks. Their technique is however restricted to single networks. 



\section{Smart grid model}
\label{sec:model}
We will first discuss the network model and then the attack model. 
\subsection{Network Model}
We consider two interdependent scale free networks, a communication network $N_A = (V_A,\alpha_A)$ and a power network $N_B = (V_B, \alpha_B)$, 
where $n_A = |V_A|$ and $n_B = |V_B|$ are the number of nodes in the communication and power networks, respectively, and $\alpha_A$ and $\alpha_B$ are the power-law coefficients. 
This implies that, $N_A$ has the power law distribution $P_A(k) \propto k^{-\alpha_A}$, which means that the fraction  of nodes
with degree $k$ is $P_A(k)$. 
Similarly, $N_B$ has the power law distribution $P_B(k) \propto k^{-\alpha_B}$. 
We assume that there are more communication nodes than power stations, which implies that $n_A >n_B$. 

The interlinks, also called  support links \cite{SBHS11} are directed edges from one network to the other. 
We assume that a communication link supports one power station and is powered by one power node, meaning that both the in-degree and out-degree of a
communication node is one. 
A power node is controlled by multiple communication node and supplies power to multiple communication nodes, 
meaning the in-degree of a power node is greater or equal to one and there is no restriction on the out-degree of a power node. 
Links are assigned randomly from the communication network $N_A$ to the power network $N_B$. 
Let $\tilde{k}_A$ denote the support degree of a node in Network $A$. 
This implies that there are $\tilde{k}_A$ nodes in $N_B$, that support a node in $N_A$. 
Let $\tilde{P}_A(\tilde{k}_A)$ denote the degree distribution of support links from $N_B$ to $N_A$.
$\tilde{P}_B(\tilde{k}_B)$ can be defined analogously. 
From the structure of the network, $\tilde{k}_A$ is equal to one for all nodes in $N_A$. 

To calculate the degree distribution $\tilde{P}_B(\tilde{k}_B)$, we note that the problem of assigning 
support links from $N_A$ to $N_B$ is equivalent to assigning $n_A$ balls randomly into $n_B$ bins. 
If $X_i$ denotes the random variable that counts the number of balls in bin $i$, 
then,  
\begin{center}
$Pr[X_i = k] = \binom{n_A}{k} \left(\frac{1}{n_B}\right)^k \left(1-\frac{1}{n_B}\right)^{n_B - k}$.
\end{center}
Thus, the degree distribution $\tilde{P}_B(\tilde{k}_B)$ follows Binomial distribution with parameters $Bin(n_A,\frac{1}{n_B})$. 

\subsection{Attack Model}
We consider targeted attack on communication network, in which the attacker attacks a node with probability proportional to the degree of the node. 
This implies that a high degree node is more prone to  attack than a low degree node. 
Targeted attacks are more likely to arise in real-world situations, as we have seen during the recent Stuxnet attack. 
Attacking the high degree node is also intuitive, since disrupting the high degree nodes result in more connections being disrupted, 
thus disrupting the network.

\begin{figure}[ht]
\centering
\includegraphics[scale=1.0]{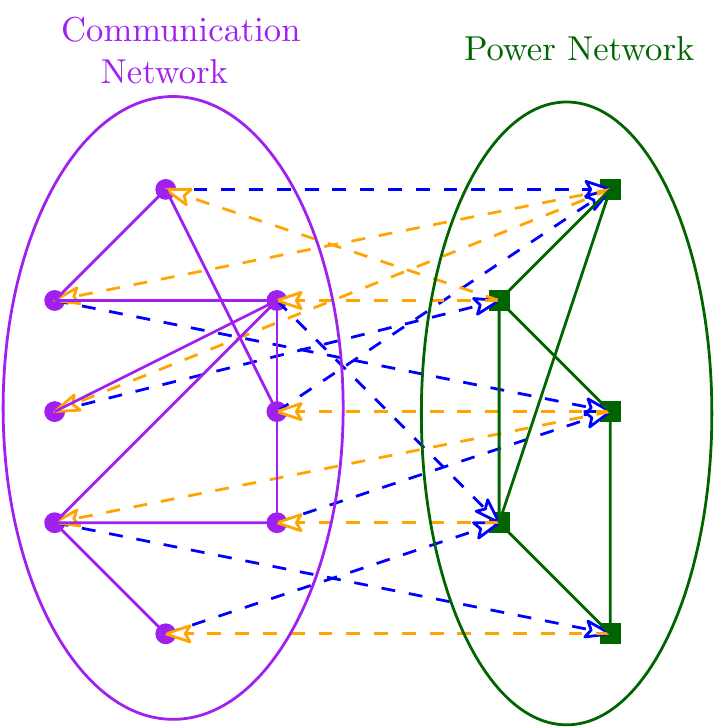}
\caption{The smart grid as an interdependent complex network.}
\label{CommPower}
\end{figure}
 
\subsection{Giant Component}
A \emph{giant component} in a graph on $n$ vertices is a maximal connected component with at least $cn$ vertices, for some constant $c$. If $c = 0.5$, this means that the giant component should have at least half of the vertices in the graph. A vertex can be deleted from the graph in two ways.

\begin{enumerate}
	\item If the vertex is attacked.
	\item If the vertex is not attacked, but all its support links on the other network has been attacked.
\end{enumerate}

Note that due to this kind of cascading failure of nodes, many more nodes will be compromised. This is different from the normal scenario, where only the attacked nodes are compromised.

\subsection{Notations}
We have used the notations in Table \ref{table:notations} throughout the paper. 
Figure \ref{CommPower} shows smart grid as an interdependent network. 

\begin{table}[t]
\label{table:notations}
\centering
\normalsize
\caption{Notations}
\label{notations1}
\scalebox{0.9}{
    \begin{tabular}{|c|p{7cm}|}
        \hline
	$N_A$ & Communication network\\
        \hline
	$N_B$ & Power network\\
        \hline
	$n_A, n_B$ & Number of nodes in $N_A$ and $N_B$ \\
        \hline
	$k$ & Degree of a node\\
	\hline
	$P_A(k)$ & Degree distribution of  communication network \\
	\hline
	$P_B(k)$ & Degree distribution of power network \\
	\hline
	$\tilde{P}_A(\tilde{k}_A)$ & Degree distribution of support degree of a node in $N_A$\\
	\hline
	$\tilde{P}_B(\tilde{k}_B)$ & Degree distribution of support degree of a node in $N_B$\\
	\hline
	$\mathcal{G}A_n$ & Giant component of $N_A$ at stage $n$\\
	\hline
	$\mathcal{G}B_n$ & Giant component of $N_B$ at stage $n$\\
	\hline
	$q_{A_k}$ & Probability of a node having excess degree $k$ (\emph{i.e.,} total degree $k+1$) in $N_A$ \\
	\hline
	$q_{B_k}$ & Probability of a node having excess degree $k$ (\emph{i.e.,} total degree $k+1$) in $N_B$ \\
	\hline
	$r_{A_n}$ & Fraction of removed nodes in $N_A$ at stage $n$, due to removal of nodes in $N_B$ at stage $n-1$ \\
	\hline
	$r_{B_n}$ & Fraction of removed nodes in $N_B$ at stage $n$, due to removal of nodes in $N_A$ at stage $n-1$ \\
	\hline
	$\mu_{A_n}$ & Fraction of functional nodes in $N_A$ at stage $n$\\
	\hline
	$\mu_{B_n}$ & Fraction of functional nodes in $N_B$ at stage $n$\\
       \hline

    \end{tabular}
    }
\end{table}

\section{Modeling cascading failure due to targeted attack on communication network}
\label{sec:analysis}
We first analyze the targeted attack on communication network and then show how the failure propagates across the interdependent networks in stages.

\subsection{Targeted attack on the communication network}
\label{subsec:target-attack-communication-node}
Let $\phi_k$ be the probability that a node $i$ of degree $k$ is not removed. 
In a targeted attack, the attacker removes a node with a probability proportional to the degree of the node. So, a high degree node is removed with higher probability. Note that, 
\begin{align*}
\phi_{k} &= 1 - \frac{deg(i)}{\sum_{v \in V_A} deg(v)} \\
&= 1-\frac{A k^{-\alpha_A}}{2m_A},
\end{align*}
Here $deg(i) = A k^{-\alpha_A}$ is the degree of node $i$ and $m_A$ is the number of edges in $N_A$.
We note that $\alpha_A = 0$ represents random removal of nodes. 

We will first calculate the size of the giant component $\mathcal{G}A_1$. 
Let $u$ denote the average probability that a node is not connected to the giant cluster via one of its neighbors. 
Consider a node of degree $k$. 
Probability that it is not connected to the giant component via any of its neighbors is $u^k$.
\begin{center}
Probability of it being in the giant component = \\
Probability that it is not attacked $\cdot$ probability that one of its neighbors is in the giant component.
\end{center}
Thus, probability of it being in the giant component is $\phi_k(1-u^k)$. 
Averaging over the degree distribution $P_A(k)$, we can calculate the fraction of nodes in the initial giant component as
\begin{equation}
\begin{split}
\mu_{A_1} &= \sum_{k=0}^{\infty}P_A(k)\phi_k(1-u^k) \\
&= \sum_{k=0}^{\infty}P_A(k)\phi_k - \sum_{k=0}^{\infty}P_A(k)\phi_k u^k \\
& = f_0(1) - f_0(u), 
\end{split}
\end{equation}
where, 
\begin{equation}
f_0(z) = \sum_{k=0}^{\infty}P_A(k)\phi_k z^k.
\end{equation}

We will now show how to calculate $u$. A node is not connected to the giant component when either of the following cases arise. 
\begin{itemize}
\item The node is attacked and thus removed, 
\item The node is present, but not connected to any node in the giant component.
\end{itemize}

Let $k$ be the excess degree of a neighboring node. The original degree of a node is one more than the excess degree, \emph{i.e.,} $k+1$ \cite{N10}. 
Probability that a neighbor is removed is $1 - \phi_{k+1}$. 
Probability that a neighbor is present, but the node itself is not present in the giant component is $\phi_{k+1}u^k$. 

Hence using \cite{N10}, $u$ can be calculated as, 
\begin{equation}
\begin{split}
u & =  \displaystyle \sum_{k=0}^{\infty} q_{A_k}(1 - \phi_{k+1} + \phi_{k+1}u^k)\\
& = 1 - f_1(1) + f_1(u), 
\end{split}
\end{equation}
where, 
\begin{equation}
f_1(z) =  \displaystyle \sum_{k=0}^{\infty} q_{A_k} \phi_{k+1}z^k. 
\end{equation}
Note that $q_{A_k}$, the probability of a node having excess degree $k$ in $N_A$ can be given by $q_{A_k} = \frac{(k+1)P_A(k+1)}{\langle k_A \rangle}$ \cite{N10}. It can be seen that, $\sum_{k=0}^{\infty} q_{A_k} = 1$. 
Substituting the value of $q_{A_{k+1}}$, the value of $f_1(z)$ can be calculated as, 
\begin{equation}
\begin{split}
f_1(z) & = \displaystyle \sum_{k=0}^{\infty} \frac{(k+1)P_A(k+1)}{\langle k_A \rangle} \phi_{k+1} z^k\\
& = \frac{1}{\langle k_A \rangle}  \displaystyle \sum_{k=1}^{\infty}k P_A(k) \phi_k z^{k-1}, 
\end{split}
\end{equation}
where, $\langle k_A \rangle$ is the average degree of nodes in $N_A$. 
We observe that, 
\begin{equation}
f_1(z) = \frac{f_0'(z)}{g_{A_0}'(1)}, 
\end{equation}
where $g_{A_0}(z)$ is the generating function, 
\begin{equation}
g_{A_0}(z) = \displaystyle \sum_{k=0}^{\infty} P_A(k)z^k.
\end{equation}

\subsection{Stage II: Effect of cascading failure on power network}
\label{subsec:stageII-powernetwork}
Due to the attack on communication nodes, the power network is affected. 
A node in the power network $N_B$ is functional, if 
a node in $N_B$ has at least one support link from $N_A$.
Initially, at stage II all nodes in $N_B$ are in the giant component. 
We consider all those nodes which are supported by nodes not in $\mathcal{GA}_1$. 
Such nodes will not remain functional because they will be cut off from the communication network. 
Probability that a node is not in the giant component $\mathcal{G}A_1$ is $1 - \mu_{A_1}$. 
Suppose, a node is supported by $\tilde{k}_B$ nodes in $N_A$. 
Probability that the $\tilde{k}_B$ neighboring nodes are not in  $\mathcal{G}A_1$ is  $(1 - \mu_{A_1})^{\tilde{k}_B}$. 
Fraction of nodes in $N_B$ disconnected due to attack on $N_A$ is given by,  
\begin{equation}
r_{B_2} = \displaystyle \sum_{\tilde{k}_B=0}^{\infty} \tilde{P}_B(\tilde{k}_B) (1 - \mu_{A_1})^{\tilde{k}_B}
\end{equation}

The fraction of nodes remaining in $N_B$ is given by $1 - r_{B_2}$. 
This is similar to the random removal of vertices. 
The fraction of nodes in the resulting giant component can be calculated by the technique in Appendix A
as 
\begin{equation}
\mu_{B_2} = (1 -r_{B_2}) (1 - g_{B_0}(u)),
\end{equation}
where,
\begin{equation} 
 u = 1 - \phi + \phi g_{B_1}(u),
\end{equation}
\begin{equation}
g_{B_0}(u) = \displaystyle \sum_{k=0}^{\infty} P_B(k) u^k 
\end{equation} and
\begin{equation}
g_{B_1}(z) = \displaystyle \sum_{k=0}^{\infty} q_{B_k}z^k.
\end{equation}

\subsection{Stage III: Cascading failure in communication  network}
\label{subsec:stageIII-commnetwork}
We will now study the effect of cascading failure in the communication network, due to the failure in power networks. 
Each node in $N_A$ is supported by only one link from the power network. 
If a node in $N_B$ fails, then the communication node it supports, also fails. 
The fraction of nodes in $N_A$ which fail due to failure of node in $N_B$ is given by, 
\begin{equation}
r_{A_3} = \sum_{\tilde{k}_A=0}^{\infty} \tilde{P}_A(\tilde{k}_A) (1-\mu_{B_2}). 
\end{equation}

We can consider that these nodes are randomly removed in $N_A$ and find the giant component resulting due to this removal of nodes. 
The fraction of nodes in the giant component which result from this random compromise is calculated as shown in Appendix A, as, 
\begin{equation}
\mu_{A_3} = (1 - r_{A_3})(1 - g_{A_0}(u)),
\end{equation}
where,
\begin{equation} 
 u = 1 - r_{A_3} + r_{A_3} g_{A_1}(u),
\end{equation}
\begin{equation}
g_{A_0}(u) = \sum_{k=0}^{\infty} P_A(k) u^k 
\end{equation} and
\begin{equation}
g_{A_1}(z) = \displaystyle \sum_{k=0}^{\infty} q_{A_k}z^k
\end{equation}.

\subsection{Stage IV: Cascading failure in power network}
\label{subsec:stageIV-powernetwork}
We now calculate the number of nodes in the power network which are connected to nodes not in the giant component in the communication network. 
The fraction of nodes which are removed because they have all their support links from the nodes not in the giant component of $N_A$, is given by, 
\begin{equation}
r_{B_4} = \displaystyle \sum_{\tilde{k}_B=0}^{\infty} \tilde{P}_B(\tilde{k}_B) (1 - \mu_{A_3})^{\tilde{k}_B}.
\end{equation}
 The giant component can be calculated as in Appendix A. 

\subsection{Giant components and steady state conditions}
We will now calculate the size of the giant component at steady state. 
Let, $r_{A_{2n-1}}$ ($n \ge 1$) be the fraction of nodes in $N_A$ that are removed due to the removal 
of nodes in $N_B$ at stage $2n-2$.  
For $n=1$, the analysis is given in Section \ref{subsec:target-attack-communication-node}. 
Then, 
\begin{equation}
r_{A_{2n-1}} = \sum_{\tilde{k_A}=0}^{\infty}(1-\mu_{B_{2n-2}})\tilde{P}_A(\tilde{k}_A). 
\end{equation}

Proceeding similarly as above, the general expression for nodes for the fraction of nodes in the giant component at the $(2n-1)$-th stage in the communication network is 
given by, 
\begin{equation}
\mu_{A_{2n-1}} = (1 - r_{A_{2n-1}}) (1 - g_{A_0}(u)),
\end{equation}
where,
\begin{equation} 
 u = 1 - \phi_{A_{2n-1}} + \phi_{A_{2n-1}} g_{A_1}(u),
\end{equation}
\begin{equation}
g_{A_0}(u) = \sum_{k=0}^{\infty} P_A(k) u^k 
\end{equation} and
\begin{equation}
g_{A_1}(z) = \displaystyle \sum_{k=0}^{\infty} q_{A_k}z^k.
\end{equation}

Similarly, let, $r_{B_{2n}}$ be the fraction of nodes in $N_B$ that are removed due to the removal
of nodes in $N_A$ at stage $2n-1$. Then, 
\begin{equation}
r_{B_{2n}} = \displaystyle \sum_{\tilde{k}_B=0}^{\infty} \tilde{P}_B(\tilde{k}_B) (1 - \mu_{A_{2n-1}})^{\tilde{k}_B}
\end{equation}
The fraction of nodes in the giant component of $N_B$ at stage $2n$ is given by, 
\begin{equation}
\mu_{B_{2n}} = (1 -r_{B_{2n}}) (1 - g_{B_0}(u)),
\end{equation}
where,
\begin{equation} 
 u = 1 - \phi + \phi g_{A_1}(u),
\end{equation}
\begin{equation}
g_{B_0}(u) = \sum_{k=0}^{\infty} P_B(k) u^k 
\end{equation} and
\begin{equation}
g_{B_1}(z) = \displaystyle \sum_{k=0}^{\infty} q_{B_k}z^k
\end{equation}.

We arrive at a steady state when, 
\begin{eqnarray}
\mu_{A_{2n-1}} & = &\mu_{A_{2n+1}} = \mu_{A_{2n+3}} = \ldots\\
\mu_{B_{2n-2}} & = & \mu_{B_{2n}} = \mu_{B_{2n+2}} = \ldots
\end{eqnarray}

It is difficult to solve these systems of equations analytically. So, we generate the smart grid using different random graph models and simulate the effect of targeted and random attacks on these graphs. The results of this study is given in the next section. 

\section{Experimental results}
\label{sec:experimental}
\subsection{Experimental Set-up}
In order to simulate a smart grid, we use the network library \emph{igraph} \cite{GRAPH} on C. 
We consider two networks: the power network and the communication network, both of which are scale-free networks. 
For each communication node, an interlink is assigned by choosing a power node at random. 
We consider three types of attack on the communication network -- targeted, random and mixed (combination of the first two). In the random attack, we choose $x$ nodes uniformly at random from all the nodes without replacement. 
In the targeted attack, we choose $x$ nodes without replacement, such that the probability of choosing a node is proportional to the degree. For mixed attacks, we select half of the nodes for targeted attack and half of the nodes for random attack.
Finally, we study the effect of compromise by running the experiment 50 times for each input $x$. 
Every time the same graphs are considered. 

\subsection{Experimental results, observations and inferences}
\begin{figure}[ht]
\centering
\includegraphics[scale=0.45]{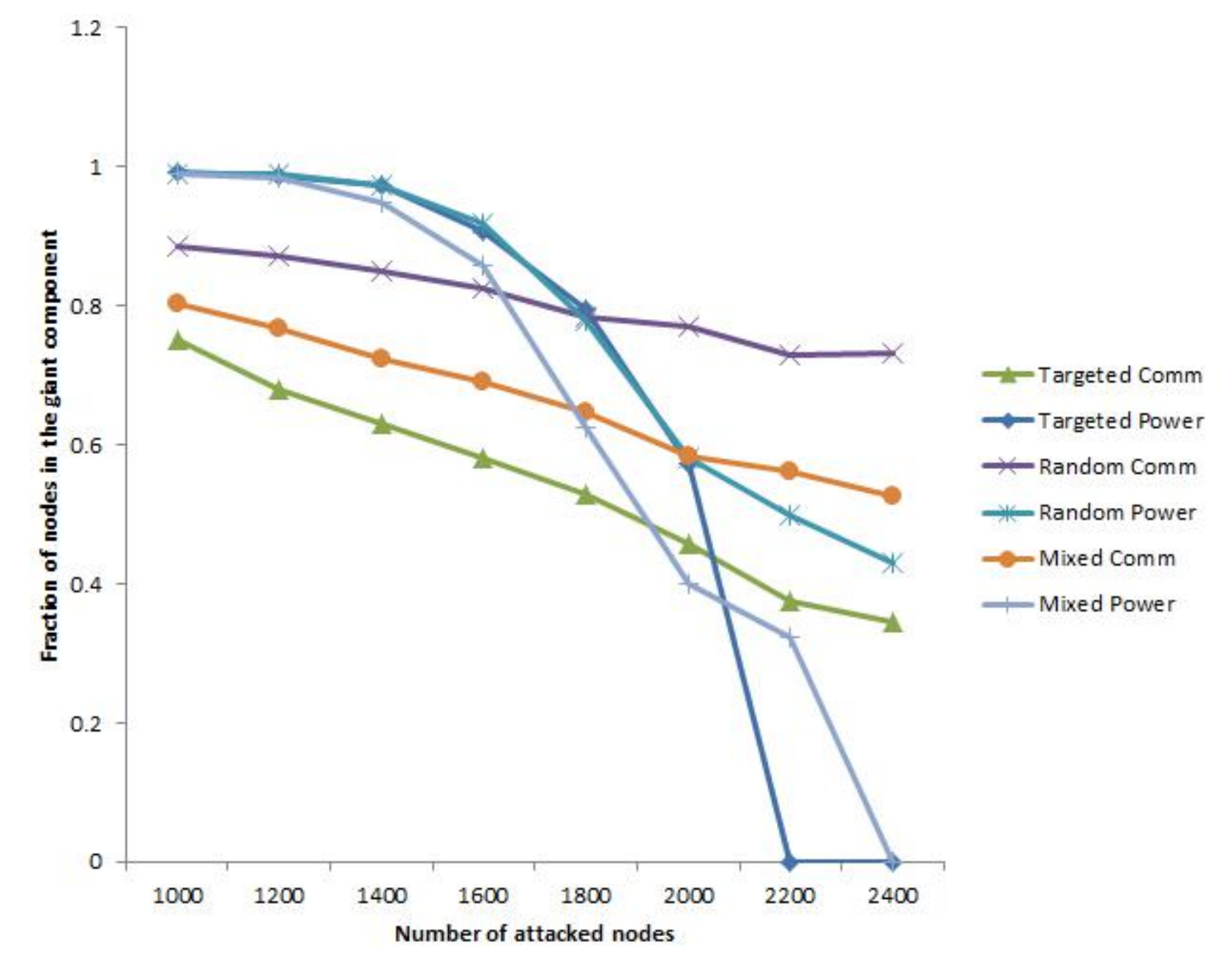}
\caption{Variation of giant component size with number of attacked nodes for targeted, random, and mixed attacks.}
\label{Smart Grid}
\end{figure}

In Figure \ref{Smart Grid}, the power network consists of 1,000 nodes and the communication network consists of 10,000 nodes. 
The communication/power network is generated as a scale-free network using a power-law degree distribution. 
We have plotted the size of the giant component (as a fraction of the size of the communication/power network) against the number of nodes attacked in the communication network. 
We observe that for a given value of the number of attacked nodes (only nodes in the communication network are attacked), 
the fraction of nodes in the giant component of the communication network is 
highest for random attacks and lowest for targeted attack. 
The corresponding fraction for mixed attacks lies somewhere in the middle.
We also see that for the same fraction of nodes compromised, the giant component of the power network disintegrates faster for targeted attacks, 
compared to random attacks. We see that on compromising 2200 nodes, there is no giant component when targeted attack occurs, whereas giant component exists, under random attacks. 
This is expected, as attacking higher degree nodes result in a faster disintegration of the network, resulting in smaller components. 

\begin{figure}[ht]
\centering
\includegraphics[scale=0.45]{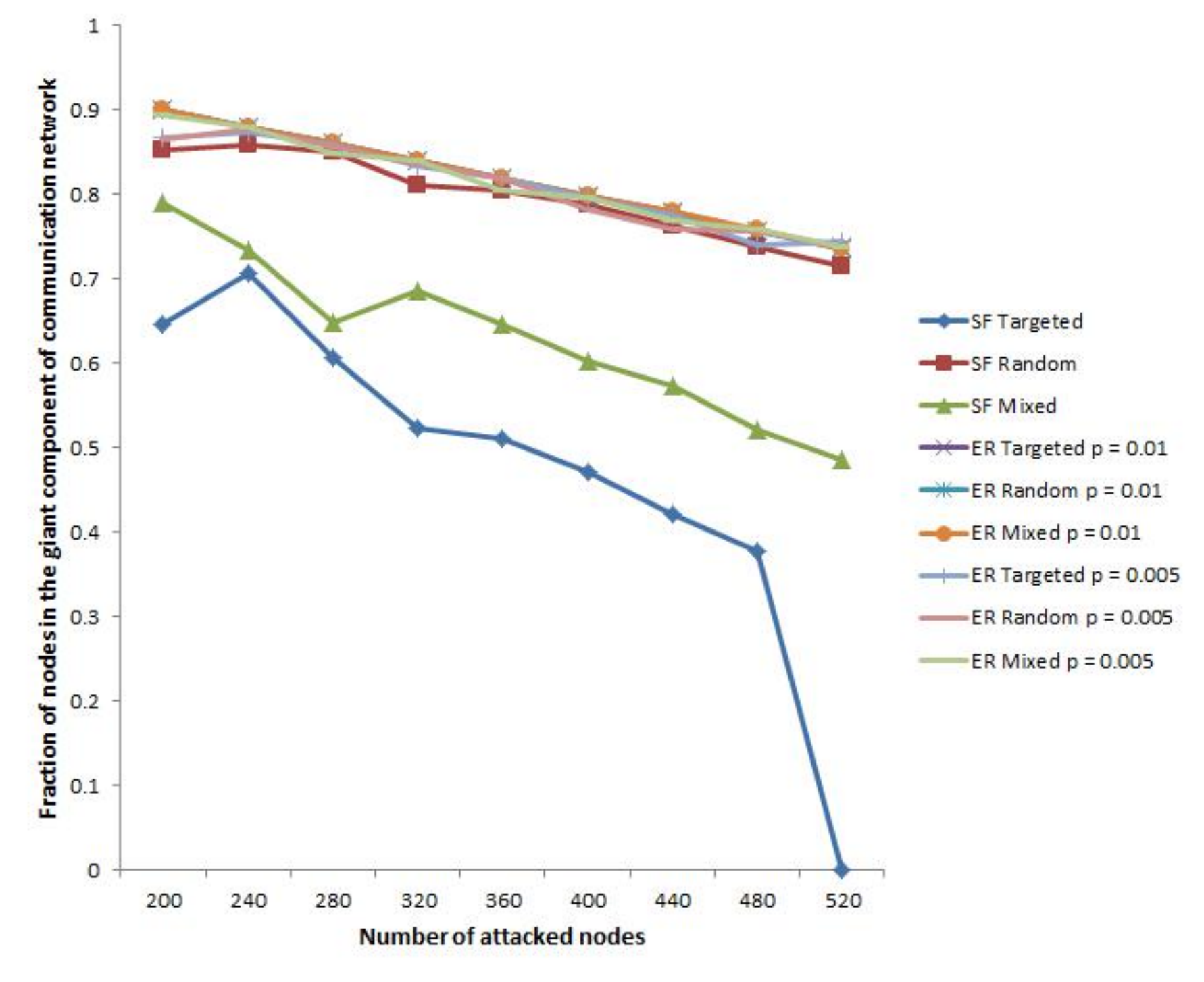}
\caption{Variation of giant component size with number of attacked nodes in the communication network for scale-free and Erdos-Renyi models.}
\label{Comm-GiantComp}
\end{figure}

\begin{figure}[ht]
\centering
\includegraphics[scale=0.45]{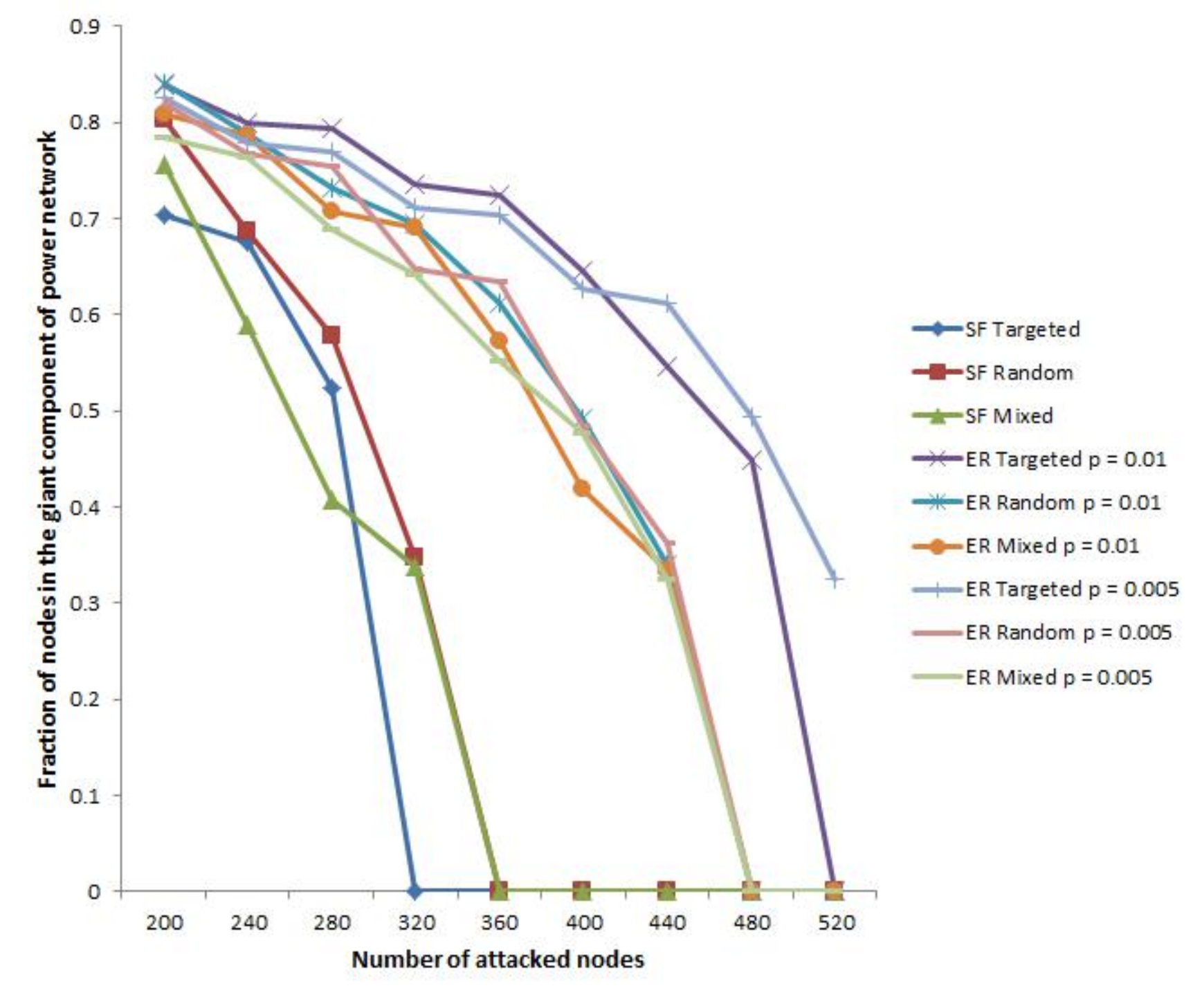}
\caption{Variation of giant component size with number of attacked nodes in the power network for scale-free and Erdos-Renyi models.}
\label{Power-GiantComp}
\end{figure}

In Figures \ref{Comm-GiantComp} and \ref{Power-GiantComp}, 
the communication network consists of 2000 nodes, whereas the power network consists of 1000 nodes. 
In Figure \ref{Comm-GiantComp}, we have plotted the size of the giant component (as a fraction of the size of the communication network) 
against the number of attacked nodes. The communication network is generated using (i) a scale-free (SF) network 
using a power-law degree distribution, (ii) the Erdos-Renyi (ER) $G(n,p)$ model with $p = 0.01$, and (iii) the Erdos-Renyi 
(ER) $G(n,p)$ model with $p = 0.005$. 
Only nodes in the communication network are attacked. 
In Figure \ref{Power-GiantComp}, we have plotted the size of the giant component 
(as a fraction of the size of the power network) against the number of attacked nodes. The power network is generated 
using the same models as mentioned above. From Figures \ref{Comm-GiantComp} and \ref{Power-GiantComp} we see that, 
for all the three types of attacks, the sizes of the giant components for ER graphs are comparable. 
The power and communication networks will be more fault tolerant to targeted attacks for Erdos-Renyi networks, compared to scale free networks. 



\section{Conclusion and future work}
\label{sec:conclusion}
We model the power and communication networks as two interdependent networks, and analyze cascading failure in smart grids for targeted attacks. 
To the best of our knowledge, this is the first work which addresses targeted attack on smart grids, which are modeled as interdependent networks. 
We have given a mathematical expression for the sizes of giant components when nodes are compromised. 
We have carried out experiments to show that a targeted attack gives an advantage to the adversary over random attacks. 
 
A challenging open problem is to obtain a closed-form solution for the size of the giant component from the mathematical analysis that we have presented.
Another important question is to present a good model of smart grids, which will be resilient to both random and targeted attacks. 
The structure of both the power and communication networks and the assignment of interlinks need to be studied. 
Thus, an important question is to find which network model and interconnection model will increase the resilience of the smart grid. 
In the future, the smart grid will be an internet of things, so a future direction of work is to propose a model, which will be resilient
to attacks and can disseminate information rapidly in the network.


\section{Acknowledgement}
This work is partially supported by DRDO sponsored project Centre of Excellence in Cryptology (CoEC), under MOC ERIP/ER/1009002/M/01/1319/788/D (R\&D) of ER\&IPR, DRDO.

\bibliographystyle{plain}
\bibliography{smartgrid}

\section*{Appendix A: Calculating giant component upon random removal of vertices}
\label{appendix-giantcomponent}
Let us consider a network $N$ have a degree distribution $P(k)$. 
Let $\phi$ be the fraction of nodes left after random removal of nodes. 
Let $u$ be the probability that a vertex is not connected to the giant component via a particular neighbor. 
If the  vertex has degree $k$, then average probability that it is not in the giant component is 
\begin{equation}
g_0(u) = \sum_k P(k) u^k, 
\end{equation}
where $g_0(z) = \sum_k P(k) z^k$, is the generating function for the degree distribution. 
Hence, the probability that a vertex belongs to a giant component is $1-g_0(u)$. 
However, the vertex itself present with a probability $\phi$. 
Thus fraction of nodes in the giant component is 
\begin{equation}
\mu_N = \phi(1-g_0(u))
\end{equation}

In order to calculate the value of $u$ we note that a node $i$ is not in the giant component if it is either removed, 
or it is present but not connected to the giant component via any of its neighbors. 
The first condition happens with probability $1-\phi$ whereas the second condition happens with probability $\phi u^k$. 
Since node $i$ can be reached following an edge, 
the value of $k$ follows the excess degree distribution 
\begin{equation}
q_k = \frac{(k+1)q_{k+1}}{\langle k \rangle},
\end{equation}
where $\langle k \rangle$ is the average degree of the network. 
Thus, averaging over this distribution we get
\begin{equation}
\begin{split}
u & = \displaystyle \sum_{k=0}^{\infty}q_k(1-\phi + \phi u^k) \\
& = 1 - \phi + \displaystyle \sum_{k=0}^{\infty} q_k u^k\\
& 1 - \phi + \phi g_1(u),
\end{split}
\end{equation}
where 
\begin{equation}
g_1(z) = \displaystyle \sum_{k=0}^{\infty} q_kz^k
\end{equation}
is the generating function for excess degree distribution.

\end{document}